# Anomalous Raman features of silicon nanowires under high pressure

Somnath Bhattacharyya[1]*, Dmitry Churochkin and Rudolph M Erasmus[1]

School of Physics and [1]DST/NRF Centre of Excellence in Strong Materials

University of the Witwatersrand, Private Bag 3, WITS 2050, Johannesburg, South Africa

The potential of silicon nanowires (SiNWs), (diameter < 10 nm) to transform into rigid bundle-like structures with distinct phonon confinement under high pressure ($\leq$ 15 GPa), instead of amorphising as per previous reports, is demonstrated using in-situ Raman spectroscopy. The newly observed splitting of the second order transverse optical (2TO) Raman mode into 2TO(L) and 2TO(W) phonon modes at $\geq$ 5 GPa establishes a highly anisotropic and mode-dependent pressure response of these SiNWs. Properties of the novel structures are superior compared to other nano-structured silicon and bulk-Si in terms of increased linear modulus, more localized phonon confinement and less anharmonicity.

___________________________________________________________________________

*email:Somnath.Bhattacharyya@wits.ac.za

In high pressure studies of semiconductor nano-crystals structural transformations of low-dimensional systems have fundamental implications for optoelectronic applications [1-7]. Such studies are particularly important to interpret the phonon confinement in silicon nanostructures (size typically less than 10 nm) [8-15]. Although bulk crystalline silicon undergoes a phase transition to a metallic phase at 12.3 GPa (hydrostatic pressure), in nano-structured silicon the shape, size and the respective distributions of the nanocrystals are thought to influence any phase transformations through confinement and surface energy effects [2,3,6,7,16]. This should be experimentally demonstrated. Silicon nano-wires (SiNWs) of diameter > 60 nm have shown structural phase transitions at high pressure coupled with an increased bulk modulus [16]. However, previous studies claimed amorphization of porous silicon structures of size less than 10 nm at ~10 GPa mostly based on first order (1TO) features [2,7,11-13]. For the Raman 1TO peak, with the decrease of particle size, phonon dispersion from the zone centre increases; however, some discrepancies between theoretical predictions and experimental data have been found [9-13]. Here we investigate the possibility of anisotropic pressure-dependent (structural) phase transformations of SiNWs (diameter < 10 nm) via *in-situ* Raman spectroscopy and the nature of phonon confinement in the modified microstructure by monitoring the comparatively weak Raman second order (2TO) features. The 2TO peak is expected to give complementary information to the 1TO peak and confinement, from which a novel structure or phase transformation, such as aligned super-structures, can be identified via monitoring of the anisotropic pressure behavior recorded at different overtones ($L$ and $W$ points) [1,15-20].

The SiNWs investigated here have been prepared by laser ablation in an oven and characterized using a number of microscopic tools to confirm the quality of the samples. The SiNWs sample, with wire diameter about 7 *nm* on average, was loaded in a membrane diamond-anvil pressure cell and pressurized in a cavity of ~200 *μm* diameter drilled in a hard stainless steel gasket. Silicone oil is used as the pressure transmitting medium. The nature of the cell is such that the pressure may be changed in fine steps by inflating or deflating a helium gas filled membrane through a fine capillary. The pressure is measured by way of the ruby-fluoresence line shift of a tiny ruby ball acting as a pressure marker loaded into the cavity alongside the sample, and which is excited by an argon-ion laser (514.5 nm) of the same set-up on which Raman measurements are recorded. Micro-Raman spectra were recorded using a Jobin-Yvon T64000 triple spectrometer operated in single spectrograph mode. Power at the sample was ~1.6 mW. The splitting of R1-R2 ruby fluorescence peaks and the Full Width at Half Maximum (FWHM) of the R1 line indicate quasi-hydrostatic conditions in the cell.

*Raman spectra at ambient pressure:* In SiNWs the prominent 1TO peak of crystalline Si (corresponding to the $\Gamma$ point in the phonon spectrum) is found to be shifted to ~510 cm$^{-1}$ from ~521 cm$^{-1}$. It also shows line broadening and an asymmetry due to phonon confinement in these nanowires [9, 11] [Fig 1a and inset]. Similar to the *first* order TO Raman peak, an asymmetry, a downshift (of up to 30 cm$^{-1}$) and the broadening of the *second* order TO phonon centered at about 950 *cm*$^{-1}$ have been observed [Fig 1b and 1c]. For c-Si, the 2TO peak consists of three symmetric components labeled by the irreducible representations $\Gamma_1$, $\Gamma_{12}$, and $\Gamma_{25}$, where $\Gamma_{12}$ makes a negligible contribution, and give rise to peaks at 922 cm$^{-1}$ {TO($X$)}, 942 cm$^{-1}$ {TO($W$)} and 979 cm$^{-1}$ {TO($L$)}. The Raman 2TO peak of the SiNWs is composed of two barely resolved peaks whose positions at ambient pressure are obtained by deconvoluting the spectra via curve fitting procedures.

*Raman spectra at high pressure:* A monotonic redshift of both the *first* and *second* order TO mode positions with the application of pressure is observed, with the slope of 2TO being twice that of 1TO [Fig. 1b, 1c and 2]. Regarding the 2$^{nd}$ order Raman scattering we note that individual information on two phonons satisfying the relationship $k_1 + k_2 = 0$ (unlike $k = 0$ in the 1$^{st}$ order scattering) can be complex to extract for c-Si (bulk) [15]. Although it is also complex for nano- or amorphous structures, the peaks attributed to these phonons can be more readily resolved by means of high pressure as shown here [Fig. 1b and 1c]. At low pressure the unresolved double peak at 930 cm$^{-1}$ is attributed to the 2TO- phonon overtone at *W* point. This unresolved splitting becomes more clearly resolved from 5 GPa, with the *L* point more clear from about 8 GPa. The *L* point intensity starts rising at ~1050 cm$^{-1}$. The *L*-point shifts with pressure but the *W* point position does not change significantly up to 15 GPa [Fig. 1b, 1c, 2a and 2b].

At 11 GPa the *L* point scattering is now clearly separated from the *W*-point; this indicates an anisotropic pressure response. At 15.6 GPa the relative intensity of the *L* and *W* points changes noticeably compared to other pressures. This observation indicates a possible structural transformation of SiNWs similar to the work reported in Ref. 16 [see Fig. 2a].

In order to investigate any possible change of microstructure, particularly in volume, the ratio of the integrated intensity of 2TO/1TO was calculated [Fig 3] [21]. This ratio at ambient pressure is found to be 0.15, which increases to 0.35 at 3 GPa. However, this value remains fairly constant in the intermediate pressure range (excluding the hysteresis), and increases further to 0.9 at 15.6 GPa. The observed increase of the ratio at high pressures could be qualitatively interpreted as an increase of the (2$^{nd}$ order) scattering due to a phase transformation which could be influenced by the change of surface area and surface energy of SiNWs. We postulate that at ~ 5 GPa the nanowires are forming some kind of aligned bundle-like structure which manifests as an increase of the 2TO/1TO peak ratio [Fig. 2a, inset and 3, inset]. Between 5 GPa and 15 GPa a relatively constant value of both the ratio and the position of the *W* point indicates that the wires (or the bundle) sustain pressure up to at least 15 GPa [Fig. 2a, inset and 3].

***Proposed model and semi-quantitative analysis:*** We start by comparing SiNWs with bulk-Si since at pressures less than 5 GPa SiNWs are randomly oriented where a bulk-like pressure response is expected [Fig. 2a, 2b inset and 3, inset]. The value of the Grüneisen parameter, [$\gamma_i = \{B(\partial\omega/\partial P)/\omega\}$] for SiNWs derived from linear fits to the data as a first order approximation, is close to unity for the 1TO peak (1.08), which is similar to bulk c-Si. The absence of any kind of splitting for the first order Raman peak in the experiment indicates that the response of the medium comprised of the randomly distributed silicon nanowires under hydrostatic pressure is similar to that of typical bulk (cubic) silicon. The dynamical equation for phonon modes $u_i$, which takes into account external pressure (P) and corresponding deformation ($\varepsilon_{lm}$) has the form $m\ddot{u} = -\left(K_{ii}^0 + \sum_{klm} K_{iklm}\varepsilon_{lm}u_k\right)$, where *m* mass and $K_{iklm}$ tensor determine any change of spring constant due to external deformation [16]. For Si, from symmetry considerations, only three independent components of the tensor $K$ [i.e. $K_{1111} = K_{2222} = K_{3333} = m\,p$, $K_{1122} = K_{2233} = K_{1133} = m\,q$, $K_{1212} = K_{2323} = m\,q$] have to be considered [see Ref. 20]. Using this comparison of bulk Si and SiNWs, we use bulk values of the compliances (S) and phenomenological parameters (i.e. $p/\omega_o^2$ and $q/\omega_o^2$), so that $S_{11} = 7.68 \times 10^{-12}$ Pa$^{-1}$, $S_{12} = -2.14 \times 10^{-12}$ Pa$^{-1}$, $p/\omega_o^2 = -1.85$ and $q/\omega_o^2 = -2.31$. Under hydrostatic pressure and after solving the corresponding secular equation we obtain the frequency shift $\Delta\omega = (p+2q)(S_{11} + 2S_{12})\sigma/2\omega_0$ [20]. Substituting $\omega_o = 520$ cm$^{-1}$ for bulk Si, and taking compression $\sigma = -p$, we obtained $\Delta\omega = 9.705 \times 10^{-12}\,\omega_0\,p$ or $\Delta\omega = 5.046\,P$, in GPa units. The rather good agreement obtained between these calculations and the experimental points support this "bulk approach" to Si nanowires [Fig. 2b]. The main difference between SiNWs and the bulk case arises from the confinement-induced red shift of $\omega_0$, which is 510 cm$^{-1}$ in our case and gives $\Delta\omega = 4.949\,P$. Thus for confined phonons in SiNWs the $\Delta\omega$ vs $P$ curve has a smaller slope than for unconfined phonons (in c-Si) with a different starting point (510 cm$^{-1}$ vs. 520 cm$^{-1}$).

To fit the pressure dependent shift of the L point we have chosen the bulk value corresponding to $\omega_0 = 979$ cm$^{-1}$, which gives $\Delta\omega = 9.501\,P$. The good agreement with experiment indicates that most of our simplifications look reasonable for the L point [Fig. 2b]. Most of the nanowires are [111]-type and we did not observe significant confinement induced shift along the [111] growth direction and $\gamma$ for L point is almost equal to the typical bulk silicon value, i.e. $\gamma = (p+2q)/6\omega_0^2 \cong 1.08$. For the W point we expect the strong confinement effect similar to that for $\Gamma$ point (at 1TO). It is well-known that the value of the red shift for 1TO is defined by the diameter of the SiNWs [11,13]. It leads to the conclusion that for both points the redshift should have a similar value, i.e. $\approx 10$ cm$^{-1}$. Then, for W point we have $\omega_0 \approx 942 - 10$ cm$^{-1}$ = 932 cm$^{-1}$, which gives $\Delta\omega = 9.045\,P$. However, for the W point we observed a significant deviation from the calculated dependence after 6 GPa. To fit the W point behavior we use the expression $\Delta\omega = 9.045\,P$ (from 0 to 6 GPa) and $\Delta\omega = 0.9045\,P$ (starting from 6 GPa) (figure 2b). A possible reason for this difference in the coefficient (i.e. the product of the compliances and $\gamma$) for the W point above and below 6 GPa by an order of magnitude compared to the bulk is now discussed. If we suppose that at low pressures (< 6 GPa) the coefficients are the same as for bulk Si, then at 6 GPa we have a crossover for the W point optical mode to a decreased frequency shift behavior. Physically it indicates that the material becomes stiffer (analogous to increase of linear or bulk modulus) under pressure for W point modes. The suggested formation of bundle-like SiNWs at higher pressure regimes explains the observed 2TO features [see Fig. 2a, inset]. Recently an almost 25% increase of the bulk modulus has been observed for high pressure phase transitions along with reduced compressibility in silicon nanowires with diameter of 60 - 80 nm [16]. It was claimed that the reason for the reduced compressibility was the unique wire-like structure of Si nanowires, in particular, the sensitivity of Young's modulus (E) to the diameter of nanowires [17]. At first glance, we can expect a less pronounced effect for $\gamma$ and compliances because Young's modulus becomes less for the smaller diameters of nanowires (less than 10 nm) we investigated [17]. The remarkable feature is that we observe a mode-dependent high pressure response (i.e. change of linear modulus and $\gamma$ in W mode), which could be explained qualitatively in the following manner. Below 6 GPa we have a typical bulk-like behavior for all modes. We have a powder of randomly distributed SiNWs, which only slightly interact with each other, i.e. most of the effects are rather well described in the frame of bulk silicon Above 6 GPa the L-mode, (i.e. the unconfined mode) still exhibits a bulk response, whereas the pressure-induced shift of the confined W-mode significantly changes [Fig. 1c and 2b]. The interaction between the rod-like structures of Si nanowires increases so that they touch and even entwine with each other. Effectively the material tends to an anisotropic bundle-like structure [see Fig. 2b]. The alignment implies further compression in directions normal to the bundle axis [111] becomes more difficult. This leads to a decrease of the compressibility in those directions whereas in directions parallel to the bundle axis it is still almost bulk-like. As a consequence the unconfined L-mode ([111] direction) is almost insensitive to such processes whereas the confined W-mode is critically sensitive [Fig. 2b, inset].

According to a previous report [16,17] the increase of B should not be significant in the present case, as the diameter of the SiNWs is less than 10 nm. So the observed change of the ratio of (B/γ) above 6 GPa for the W point should be mostly accounted for by the decrease of the value of γ by at least a factor of 10. Below 6 GPa the random orientation of the nanowires gives rise to a strong anharmonic nature of the vibration (for all modes) [20]. However, above 6 GPa there is a significant decrease of anharmonicity (of the phonon vibration) of the W-mode which is consistent with the proposed model of aligned and bundle-like structures of nanowires at high pressure [Fig. 2b, inset]. At high pressure these wires acting as a bundle can sustain pressure and the pressure response can be explained by the change in the phonon spectra, which may arise from the change of the nature of confinement [22]. In order to quantify the change of structure we plot the single parameter $\partial\omega/\partial P$ as a function of P which clearly shows a discontinuity in the curve at about 6 GPa which strongly suggests a structural change of the materials at this point as depicted as inset of Fig. 2a & 2b.

The present material behaves in a very different way from porous silicon (PS), where porosity (air gap between nanocrystals) can impose a strong confinement in these structures. These (PS) nano-crystals cannot form a single unit under pressure and rather undergo amorphization, unlike SiNWs. The proposed alignment of the SiNWs can be explained to first order by a volume reduction, mechanisms for precise control of the alignment remaining unknown at present. Any process that offers nano-scale control of phonon confinement related properties in the bundle structures is advantageous for potential nano-optoelectronics applications. Photoluminescence properties of these SiNWs bundles are expected to be improved significantly under high pressure [see Ref. 7], which is a prerequisite for fabricating nano-optoelectronic (e.g. pressure sensor) devices which can be easily achieved by pressurizing SiNWs in a sample cell.

*Acknowledgments:* SB is very grateful to Professor G. R. Hearne for his assistance with the experiment and to Professor A.G. Every for valuable discussions.

**Figure captions:**

**Fig. 1(a)** Variation of 1TO Raman peak of SiNWs with pressure. **Inset:** 1TO peak upon decompression. **(b)** Variation of Raman 2TO peak with increasing pressure and **(c)** 2TO peaks upon decompression. This serves to depict the evolution of the separation of the two peaks of the 2TO mode arising from W and L points in silicon. Raman spectrum for Si (100) (bulk) at ambient pressure is included.

**Fig. 2(a)** Calculated parameter $\partial\omega/\partial P$ shows a transition similar to a phase transition at about 6 GPa. **Inset:** Schematic diagram of the formation of proposed superstructure of Si nanowires with increasing pressure. Left panel illustrates randomly oriented nanowires at ambient pressure. Right panel shows aligned bundle structure.

**Fig. 2(b)** Calculated shift of $\Gamma$ point (1TO), W and L points (2TO) with pressure (solid lines) to fit the experimental data (open symbols). **Inset:** Schematic diagram showing the proposed mode-dependent pressure response model, confinement and a crossover from anharmonic to harmonic vibration in the SiNWs at high pressure.

**Fig. 3** Ratio of 2TO to 1TO peak intensity with pressure (arrows distinguish the compression and decompressions series). Green arrows (broken lines) guide the eye for the suggested trend of the ratio with pressure. Error bars represent a standard deviation of the data. **Inset** shows scanning electron micrograph of randomly oriented SiNWs.

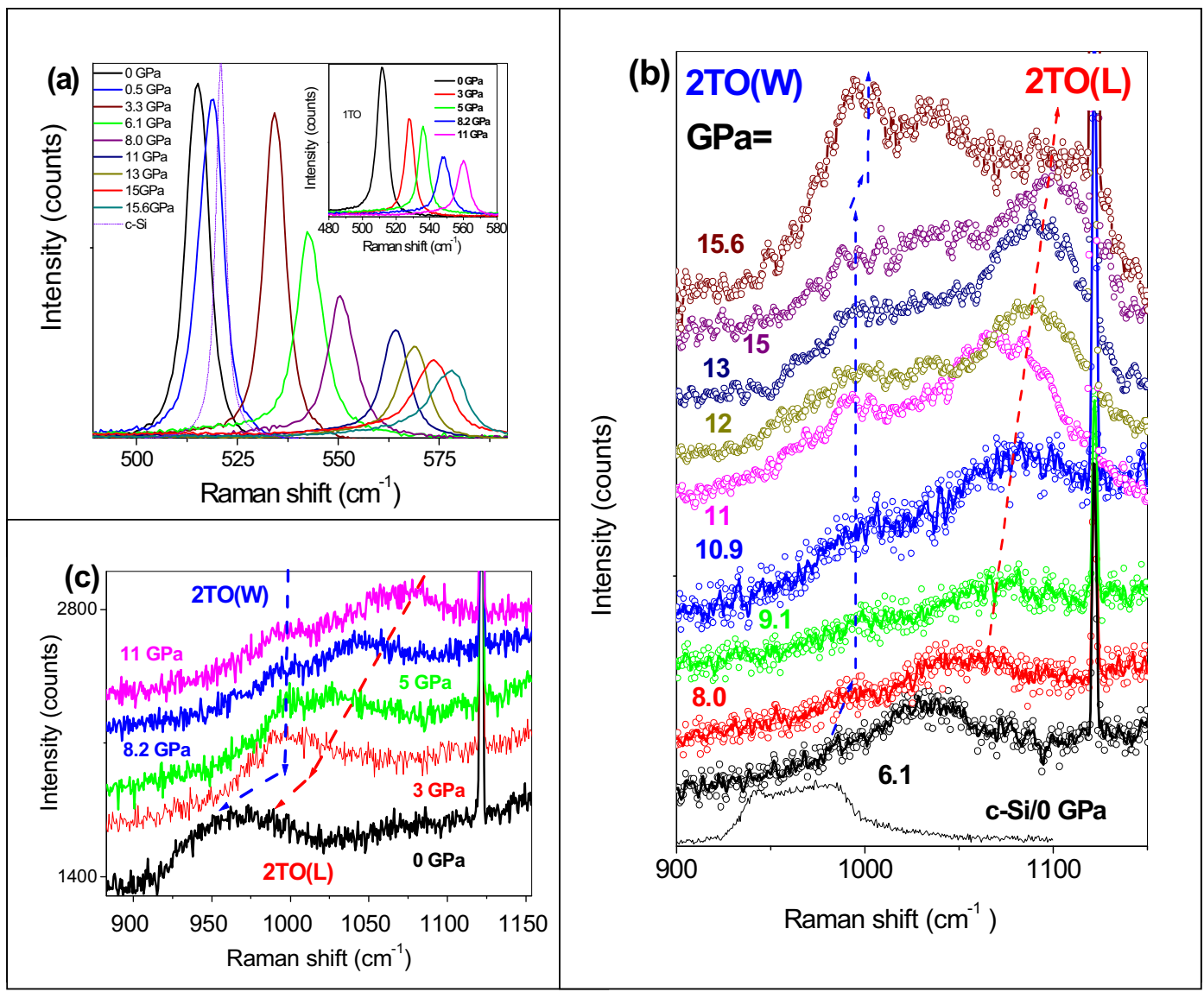

**Fig. 1 (a) (b) and (c)**

Bhattacharyya *et. al.*

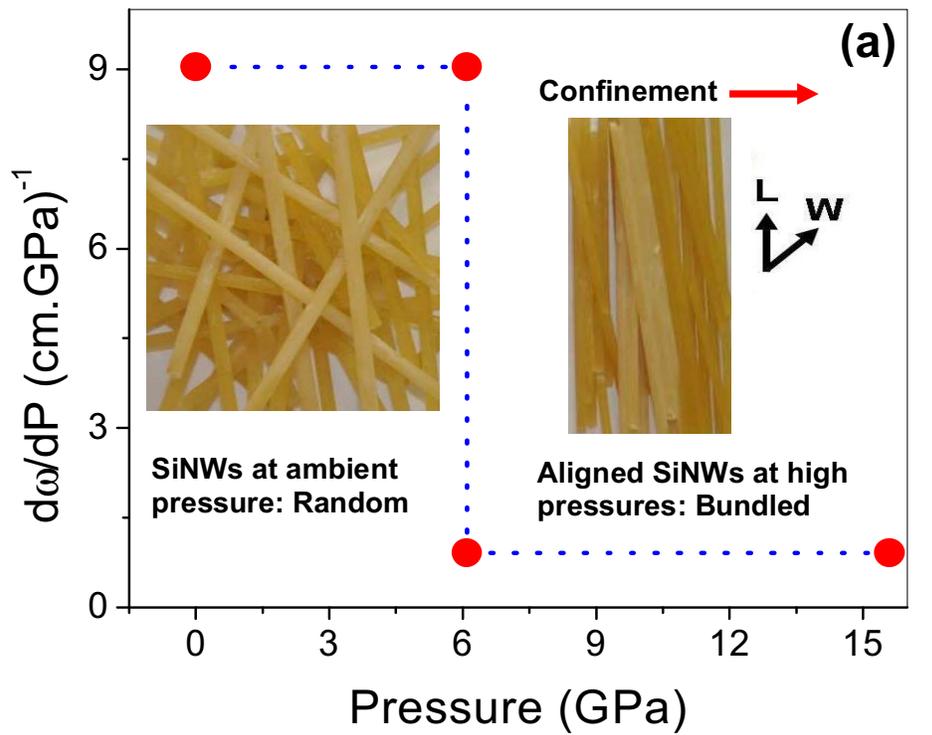

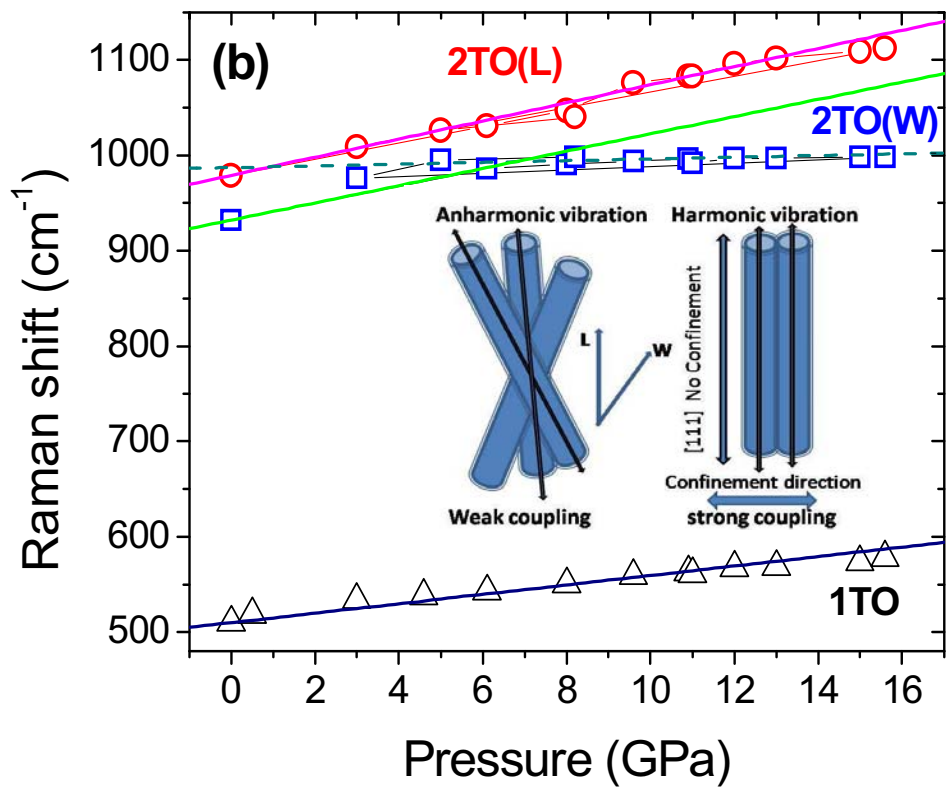

Fig. 2 (a) and (b)  Bhattacharyya *et al.*

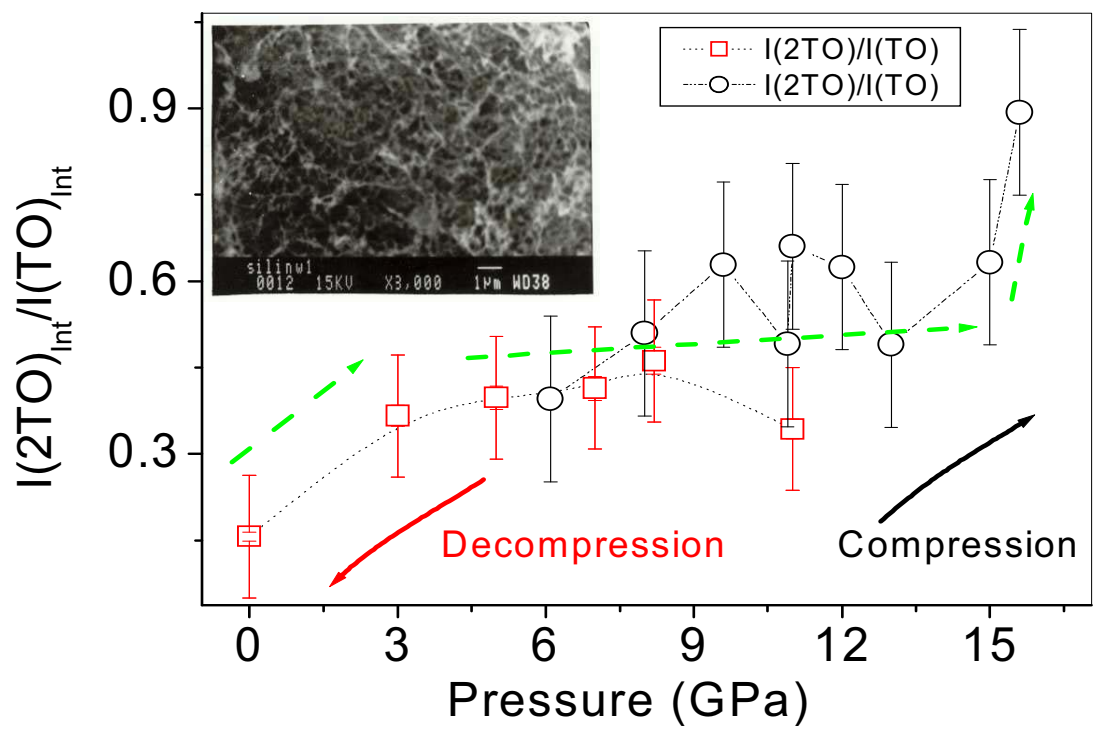

**Fig. 3**

Bhattacharyya *et al.*